\documentstyle[preprint,eqsecnum,aps]{revtex}

\begin{document}
\title{Five-dimensional scenario with a fluctuating three-brane: an stochastic approach to gravity}
\author{I. Quiros\thanks{israel@uclv.etecsa.cu}}
\address{Dpto. Fisica. Universidad Central de Las Villas. Santa Clara CP 54830. Villa
Clara. Cuba}
\date{\today}
\maketitle

\begin{abstract}

 A five-dimensional scenario with a non compact extra dimension of infinite extent is studied, in which a single three-brane is affected by small Gaussian fluctuations in the extra dimension. The average magnitude of the fluctuations is of order of the electro-weak length scale ($\sigma\sim m_{EW}^{-1}$). The model provides an stochastic approach to gravity that accounts for an alternative resolution of the mass hierarchy problem. The cosmological constant problem can be suitably treated as well. Surprisingly the Mach's principle finds a place in the model. It is argued that the Mach's principle, the mass hierarchy and the cosmological constant problem, are different aspects of a same property of gravity in this model: its stochastic character. Thin-brane scenarios are recovered in the "no-fluctuations" limit ($\sigma\rightarrow 0$).

\end{abstract}


\section{Introduction}

Over the last two years space-time models with large extra dimensions have become very popular\cite{arkani-hamed,randall,massh,sundrum}. These scenarios start from the assumption that ordinary matter fields, in particular standard model (SM) particles, are confined to a three-brane (four space-time dimensions) embedded in a higher-dimensional bulk space-time. Brane scenarios account, in particular, for successful resolutions of the mass hierarchy problem\cite{arkani-hamed,massh,sundrum,lykken}. Even if none of these scenarios with large extra dimensions correctly describe our present universe, their potential for describing the early universe is obvious since there is strong evidence that the early universe underwent a phase where it was five-dimensional\cite{witten}.

Among the models with large extra dimensions, the Randall-Sundrum (RS) set-up is one of the most appealing scenarios\cite{randall,massh,sundrum}. This scenario relies on the existence of just one extra dimension. There are two three-branes with equal but opposite tensions embedded in the five-dimensional bulk space-time. The graviton is confined to one of these branes as a result of a specific ansatz for the line-element

\begin{equation}
ds^2=e^{-|y|}\eta_{nm}dx^n dx^m + \frac{1}{4k^2}dy^2,
\end{equation}
where the Minkowski metric $\eta_{ab}$ is multiplied by a warp factor. Orbifold symmetry has been imposed upon the extra $y$-coordinate ($y\rightarrow -y$) that takes values in the interval $y\in [0,2k\pi]$. The line-element (1.1) is consistent, besides, with Poincare invariance. In general, Randall-Sundrum-type models are based on the following action\cite{massh,sundrum}

\begin{equation}
S_5=-\int d^4x dy \sqrt{-G}\{\frac{R_5}{2k_5^2}+2\Lambda\}+\sum_\alpha\int d^4x \sqrt{-g_\alpha}{\cal L}_\alpha,
\end{equation}
where $R_5$ is the Ricci scalar made out of the five-dimensional metric $G_{AB}$ , $k_5^2$ is the five-dimensional Planck scale, $\Lambda$ is the bulk cosmological constant and $\alpha$ labels the different branes in the bulk. The four-dimensional metrics induced on the branes at $y=y_\alpha$ are $g_{ab}^\alpha=G_{ab}(x,y=y_\alpha)$, where $x\equiv x^a (a=\overline{0,3})$, accounts for the set of usual four-dimensional space-time coordinates. The Lagrangian ${\cal L}_\alpha={\cal L}_\alpha^{matter}+V_\alpha$, includes both the matter degrees of freedom and the brane tensions. The following field equations can be derived from the action (1.2)\footnotemark\footnotetext{For simplicity we are not considering the dynamics of the matter fields on the branes}

\begin{equation}
\sqrt{-G}(R_{AB}-\frac{1}{2}G_{AB}R_5)=-k_5^2\sum_\alpha \sqrt{-g_\alpha}g_{mn}^\alpha\delta^m_A\delta^n_B{\cal L}_\alpha \delta(y-y_\alpha)+2k_5^2\sqrt{-G}\Lambda G_{AB}.
\end{equation}

The presence of Dirac delta functions in the field equation (1.3) yields that, although the metric is required to be continuous across the branes, its derivatives with respect to $y$ may be discontinuous at $y=y_\alpha$. In particular, delta functions arise in the second derivatives of the metric with respect to $y$. This implies, in turn, that the field equations, which contain up to second order derivatives of the metric, are undefined at the brane locations $y=y_\alpha$, unless appropriate junction conditions are imposed\cite{deffayet,kanti}. When one deals with Friedmann-Robertson-Walker (FRW) cosmology on the branes, imposition of such junction conditions yields that a new Friedmann-type equation is obtained\cite{deffayet}. It is fundamentally different from the standard cosmological Friedmann equation in that the square of the Hubble parameter depends quadratically on the cosmological energy density and, besides, it contains the five-dimensional Newton's constant (the fundamental Planck mass scale of the higher-dimensional theory) instead of the four-dimensional Newton's constant. Departure from standard Friedmann behavior entails troublesome observational difficulties for RS-type scenarios.

The above cosmological difficulties are associated with the assumption of infinitely thin branes and empty bulks in these scenarios\cite{kanti}. Rigorously, physical branes will have a width in the fifth dimension\cite{deffayet}. The thickness of the wall, at present, should be given (at most) by the electro-weak mass scale $m_{EW}\sim 10^3 GeV$ since electro-weak interactions have been probed at distances $\sim m_{EW}^{-1}$\cite{arkani-hamed}. Besides, in sufficiently hard collisions of energy $E_{esc}>m_{EW}$, the standard model particles may acquire momentum in the $y$-direction and escape from the wall\cite{arkani-hamed}. This means that, in general, one should consider a five-dimensional stress-energy tensor for the matter degrees of freedom that can propagate in the bulk. In Ref.\cite{kanti} Friedmann cosmology on a brane embedded in a five-dimensional bulk was considered by incorporating both ingredients: thick wall and non-null bulk component of the stress-energy tensor.

In the present letter we propose an alternative scenario in which a single three-brane develops small Gaussian fluctuations about some equilibrium configuration in a five-dimensional space-time of infinite extent in the extra dimension. The effect of the brane fluctuations is to generate an effective five-dimensional subspace in the bulk. The effective width of the resulting subspace in the extra dimension is set by the magnitude of the fluctuations, taken here to be $\sigma\sim m_{EW}^{-1}$. In consequence we have an effective five-dimensional domain of thickness $\sigma$ embedded in the bulk, with a non null (effective) bulk component of the matter stress-energy tensor. In this sense our scenario may represent an alternative to the model of Ref.\cite{kanti} allowing for vanishing of the cosmological difficulties arising in (thin-brane) RS-type scenarios. However, in this letter we shall consider, for simplicity, that brane fluctuations are stabilized in the sense that, neither the width of the distribution nor its center depend on time. In other words, we are considering a static configuration in the five-dimensional space-time. Consequently, it does not seem to be a good framework for treating the cosmological issue. Nevertheless, the approach can give some insight into cosmological behavior by extrapolating the results obtained for stabilized brane fluctuations back into the past. A detailed cosmological study of the present scenario (with non stabilized fluctuations) will be left for future work.

Here we shall focus in the stochastic nature of gravity that is implicit in our model. It provides an (stochastic) approach to the description of the universe that, in some sense, resembles that of Ref.\cite{linde}. In this context we propose an alternative (stochastic) resolution of the mass hierarchy problem. Recall that this is one of the main achievements of thin-brane scenarios. It is remarkable, besides, that the cosmological constant problem can be suitably studied in the context of the present (stochastic) approach to gravity. It is worthy of mention, also, that the Mach's principle finds a place in this model. All of them: The mass hierarchy, the cosmological constant problem and the Mach's principle, seem to be different aspects of the stochastic nature of gravity. Thin-brane scenarios are recovered in the "no-fluctuations" limit ($\sigma\rightarrow 0$).

The letter has been planed as follows. In section II we explain the details of the model. An action, that represents a generalization of RS-type one to include the effect of brane fluctuations, is presented and the corresponding field equations are obtained. The conservation equations are also given. A Poincare invariant solution is studied in section III. This solution provides an alternative resolution of the mass hierarchy problem to be studied in section IV. It is due to a remarkable circumstance that, Gaussian fluctuations of the brane induce Gaussian fluctuations of the metric in the extra dimension, but with a different width and a different location of its center. The mass hierarchy problem is first solved "in average" and then the stochastic resolution is provided. In section V we study in detail the stochastic character of four-dimensional gravity that emerges from our model. Special emphasis is made in the cosmological constant problem. Surprisingly the Mach's principle finds place in the model and is briefly discussed in section VI. Conclusions are given in section VII.

\section{The model}

The present model is based on the assumption that the five-dimensional bulk can be considered as a macroscopic isolated thermodynamic system in statistical equilibrium. In this state the system is characterized by some equilibrium (absolute) temperature ${\it T}_0$. Each three-brane in the bulk can be considered as quasi-isolated subsystem immersed in a thermostat (the bulk) of temperature ${\it T}_0$. This last assumption holds true since we are considering that all (SM) particles (with the exception of the graviton) are confined to one such brane from which they can not escape. In other words, we are considering particles with energies below the threshold $E_{esc}\sim m_{EW}$. We shall further assume that fluctuations may arise in one or several quasi-isolated subsystems (branes), in such a way that the thermostat (the bulk) experiences a quasi-static process without breaking its equilibrium state. The state of a given subsystem will be characterized by some external parameter, for instance, the coordinate $y$ of the brane in the extra dimension. When the quasi-isolated subsystem fluctuates about its equilibrium state, the external parameter $y$ fluctuates about its equilibrium value $y_\alpha$ - the position of the brane in the bulk.

In this letter we shall consider small fluctuations (of order $\sim m_{EW}^{-1}$) of the branes about their "equilibrium" positions in the bulk. For simplicity we shall assume a single fluctuating brane in the five-dimensional space-time. Generalization to multiple fluctuating branes is straightforward. 

If one further considers that the fluctuations of the brane in the bulk may be viewed as a displacement of a quasi-isolated subsystem in an external field of force, then, the probability to find the subsystem in a state characterized by a non-equilibrium value of the external parameter in the interval $[y,y+dy]$ is given by the Boltzmann expression

\begin{equation}
dw=const.\exp[-\frac{U(y)-U(0)}{b{\it T}_0}]\;dy,
\end{equation}
where $U(y)$ is the potential energy of the subsystem in the external field, $U(0)$ represents its potential energy in the equilibrium state (position $y=0$ in the extra dimension), $b$ is the Boltzmann constant and, as before, ${\it T}_0$ is the absolute temperature of the thermostat (the bulk). By a proper choice of the origin for the potential energy one can set $U(0)=0$. We are considering that the brane fluctuates about an "equilibrium" hypersurface located at $y=0$. Small fluctuations of the system about its equilibrium state cause that the potential function $U(y)$ can be expanded in Taylor series about $y=0$ (we consider the first non-vanishing terms in the expansion)

\begin{equation}
U(y)=U'(0)\;y+\frac{1}{2}U''(0)\;y^2+\ldots,
\end{equation}
where the prime denotes derivative in respect to the extra $y$-coordinate. In the equilibrium state the potential energy $U(0)$ is a minimum, meaning that $U'(0)=0$ and $q_0\equiv U''(0)>0$. In consequence, Eq. (2.1) can be written as a Gaussian probability distribution

\begin{equation}
dw=const. \exp[-\frac{q_0 y^2}{2b{\it T}_0}]\;dy.
\end{equation}

The magnitude of the constant $q_0$ depends on the nature of the "fifth-force" originating the displacement of the subsystem from its equilibrium configuration. The average value of the fluctuation of the external parameter $y$ (its dispersion) is given through the following expression

\begin{equation}
\sigma^2=\overline{y^2}=\int dy\;y^2\;w=\frac{b{\it T}_0}{q_0},
\end{equation}
where the following Gaussian distribution function has been used

\begin{equation}
w=\frac{1}{\sqrt{2\pi}\sigma}\exp[-\frac{y^2}{2\sigma^2}].
\end{equation}

Therefore with decreasing of the absolute temperature the magnitude of the fluctuations gets decreased and, in the "small temperature" limit, fluctuations cease. In this limit we have just a RS-type scenario with a single thin brane embedded in the bulk. In our model we consider $\sigma\sim m_{EW}^{-1}$ so the absolute temperature ${\it T}_0\sim m_{EW}^{-2}$.

The distribution function (2.5) is the basic piece of our approach with fluctuating branes. A generalization of the RS-type action (1.2), to include the effect of fluctuations of the branes, can be written in the form

\begin{equation}
S_5=-\int d^4x dy \sqrt{-G}\{\frac{R_5}{2k_5^2}+2\Lambda\}+\sum_\alpha\int d^4x dy \sqrt{-G}{\cal L}_\alpha w_\alpha,
\end{equation}
where $w_\alpha$ are the distribution functions (given through expressions of the kind (2.5)) with centers at $y=y_\alpha$. They account for fluctuations of the branes located at these positions in the extra coordinate. Standard RS-type (thin-brane) scenarios are recovered in the "no-fluctuations" limit $\sigma_\alpha\rightarrow 0$. In this limit $w_\alpha\rightarrow\delta(y-y_\alpha)$ and the action (2.6) transforms back into action (1.2).

As stated before, we shall study here a five-dimensional scenario with a single fluctuating brane so, in what follows, we shall drop the index $\alpha$. Starting with the action (2.6) one can derive the following field equations

\begin{equation}
R_{AB}-\frac{1}{2}G_{AB}R_5=k_5^2(T_{AB}-VG_{AB})w+2k_5^2\Lambda G_{AB},
\end{equation}
where we have defined $T_{AB}=\frac{2}{\sqrt{-G}}\frac{\delta(\sqrt{-G}{\cal L}^{matter})}{\delta G^{AB}}$. The explicit form of Eq. (2.7) shows that it is defined everywhere in the extra dimension. In other words, unlike Eq. (1.3) that is undefined at the hypersurfaces (slices of the five-dimensional bulk) located at $y=y_\alpha$, singular hypersurfaces do not arise in the scenario with fluctuating branes.

The "conservation" equations compatible with Eq. (2.7) are the following

\begin{equation}
T^{AN}_{\;\;\; ;N}w+(T^{AN}-V G^{AN})w_{,N}=0,
\end{equation}
or, if we consider stabilized fluctuations,

\begin{equation}
T^{AN}_{\;\;\; ;N}=0.
\end{equation}

This last equation implies, in turn, that usual (general relativity) conservation law $T^{an}_{\;\;\; ;n}=0$ holds true in this case.

\section{Poincare invariant solution}

In this section we shall derive a Poincare invariant solution to our model that will be used as the basis for further discussion. For this purpose we start with the following, Poincare invariant ansatz for the line-element\cite{sundrum}

\begin{equation}
ds^2=e^{-\psi(y)}\eta_{nm}dx^n dx^m+\frac{1}{4k^2}dy^2,
\end{equation}
where the Minkowski metric $\eta_{ab}$ is multiplied by a warp factor and $\frac{1}{4k^2}$ is some constant parameter that sets the scale for proper distance measurements along the extra dimension. Taking into account the ansatz (3.1) for the line-element the field equations (2.7) can be written explicitly as follows

\begin{equation}
(-\psi''+\psi'^2)\eta_{ab}=\frac{k_5^2}{6k^2}(T_{ab}^4-V\eta_{ab})w+\frac{k_5^2}{3k^2}\Lambda\eta_{ab},
\end{equation}

\begin{equation}
\psi'^2=\frac{2k_5^2}{3}(T_{44}-\frac{V}{4k^2})w+\frac{k_5^2}{3k^2}\Lambda,
\end{equation}
where $T_{ab}^4$ is the four-dimensional stress-energy tensor of the matter degrees of freedom living on the brane, given in terms of the Minkowski metric $\eta_{ab}$. In other words, $T_{ab}^4=e^\psi T_{ab}^5$, where $T_{ab}^5$ is the four-dimensional part of the five-dimensional stress-energy tensor $T_{AB}$, while $T_{44}$ is its bulk component. It is non zero since fluctuations of the brane generate an effective five-dimensional subspace in the bulk with non-null (effective) energy content.

Fluctuations of the brane about the "equilibrium" hypersurface $y=0$, induce fluctuations of the metric $G_{AB}(x,y)$ in the extra dimension. The following expansion of the exponent in the warp factor in Eq. (3.1) takes place

\begin{equation}
\psi(y)=\psi'(0)\; y+\frac{\psi''(0)}{2}\;y^2+\ldots,
\end{equation}
where we have taken $\psi(0)=0$ and, as before, we have considered up to second term in the expansion about the equilibrium configuration at $y=0$. The constant factors $a_0=\psi'(0)$ and $b_0=\psi''(0)$ in the expansion (3.4) can be found with the help of the field equations (3.2) and (3.3) evaluated at $y=0$,

\begin{equation}
a_0=(\frac{2k_5^2}{3\sqrt{2\pi}})^\frac{1}{2}\sigma^{-\frac{1}{2}}(T_{44}-\frac{V}{4k^2}+\sqrt{\frac{\pi}{2}}\frac{\sigma}{k^2}\Lambda)^\frac{1}{2},
\end{equation}

\begin{equation}
b_0=\frac{2k_5^2}{3\sqrt{2\pi}}\sigma^{-1}(T_{44}-\frac{1}{16k^2}T^4),
\end{equation}
where $T^4\equiv\eta^{nm}T_{nm}^4$ is the trace of the four-dimensional stress-energy tensor. The line-element (3.1) can then be written as

\begin{equation}
ds^2={\bar w}{\bar \eta}_{nm}dx^n dx^m+\frac{1}{4k^2}dy^2,
\end{equation}
where we have considered the rescaled Minkowski metric ${\bar \eta}_{ab}=\sqrt{\frac{2\pi}{b_0}}e^\frac{a_0^2}{2b_0}\eta_{ab}$, and the distribution function for fluctuations of the metric in the extra dimension

\begin{equation}
{\bar w}=\frac{{\bar\sigma^{-1}}}{\sqrt{2\pi}}\exp[-\frac{(y-{\bar y})^2}{2{\bar\sigma}^2}],
\end{equation}
with dispersion ${\bar\sigma}^2=\frac{1}{b_0}$ and the "shifted" center of the distribution placed at ${\bar y}=-\frac{a_0}{b_0}$. We see that the probability to find the system in a configuration with non-equilibrium metric is given by a Gaussian distribution like $w$ in Eq. (2.5), but with a different characteristic width ${\bar\sigma}$ and a shifted equilibrium value ${\bar y}$ of the external parameter. It is a remarkable result. In the limit $\sigma\rightarrow 0$, ${\bar\sigma}\rightarrow 0$ (no fluctuations in the extra dimension), we recover a RS-type scenario with two three-branes, one located at $y=0$ where the matter degrees of freedom are trapped, and the other at $y={\bar y}$ in which is confined the bound graviton. Recall that, in this limit, $w\rightarrow\delta(y)$ while ${\bar w}\rightarrow\delta(y-{\bar y})$.

Both, the characteristic width and the center of the distribution (3.8), depend on the matter content of the brane and on the average value of the fluctuations of the external parameter through expressions that are derivable from (3.5) and (3.6):

\begin{equation}
{\bar\sigma}^2=\frac{3\sqrt{2\pi}\sigma}{2k_5^2(T_{44}-\frac{1}{16k^2}T^4)},
\end{equation}

\begin{equation}
{\bar y}=-(\frac{3\sqrt{2\pi}\sigma}{2k_5^2})^\frac{1}{2}\frac{(T_{44}-\frac{V}{4k^2}+\sqrt{\frac{\pi}{2}}\frac{\sigma}{k^2}\Lambda)^\frac{1}{2}}{T_{44}-\frac{1}{16k^2}T^4},
\end{equation}
leading to the following relationship

\begin{equation}
\frac{{\bar y}^2}{{\bar\sigma}^2}=\frac{T_{44}-\frac{V}{4k^2}+\sqrt{\frac{\pi}{2}}\frac{\sigma}{k^2}\Lambda}{T_{44}-\frac{1}{16k^2}T^4}.
\end{equation}

Expressions (3.9-3.11) are not valid for a five-dimensional vacuum-like fluid matter content since, in this case, $T_{44}=\frac{1}{16k^2}T^4$ and these equations are undefined. Since both ${\bar\sigma}^2$ and $\frac{{\bar y}^2}{{\bar\sigma}^2}$ can not be negative then, the following constrains on the matter content arise

\begin{equation}
T^4<16k^2T_{44},
\end{equation}
and

\begin{equation}
4k^2T_{44}\geq V-4\sqrt{\frac{\pi}{2}}\sigma\Lambda.
\end{equation}

From equations (3.9) and (3.10) one sees that, increasing of the fluctuations of the brane about the equilibrium configuration at $y=0$ yields that, fluctuations of the metric about the hypersurface with position $y={\bar y}$ in the extra dimension, increase quadratically. Besides, the position ${\bar y}$, depends on the brane tension , the bulk cosmological constant and the absolute temperature $T_0$ through $\sigma$.

\section{Mass hierarchy}

We now elaborate on these results to generate the correct mass hierarchy. Throughout the extent of this section we shall follow (practically) step by step the procedure used in Ref.\cite{massh}, to explore an alternative resolution of the mass hierarchy problem within the context of our model with a fluctuating brane. We hope the results obtained justify repeating of that procedure.

\subsection{The problem in average}

We first identify the massless gravitational fluctuations about Minkowski space

\begin{equation}
ds^2={\bar w}({\bar \eta}_{nm}+{\bar h}_{nm}(x))dx^n dx^m+\frac{1}{4k^2}dy^2,
\end{equation}
where ${\bar h}_{ab}(x)$ are tensor fluctuations about Minkowski metric (the physical graviton of the four-dimensional effective theory). If we introduce the perturbed Minkowski metric

\begin{equation}
{\bar g}_{ab}={\bar \eta}_{ab}+{\bar h}_{ab},
\end{equation}
and substitute Eq. (4.1) into the action (2.6), then the pure curvature term in (2.6) can be written in the form of an action for a four-dimensional effective theory

\begin{equation}
S_{eff}\supset\int d^4x\;\int_{-\infty}^\infty dy\frac{{\bar w}}{4kk_5^2}\sqrt{-\bar g}{\bar R},
\end{equation}
where ${\bar R}$ is the four-dimensional Ricci scalar made out of the metric ${\bar g}_{ab}$. In Eq. (4.3) integration in $y$ is to be performed from $-\infty$ to $+\infty$ since the extra dimension is assumed to be infinite in extent. Explicit integration in $y$ in Eq. (4.3) yields

\begin{equation}
G_N^{-1}=\frac{4\pi}{kk_5^2}\int_{-\infty}^\infty dy{\bar w}=\frac{4\pi}{kk_5^2},
\end{equation}
where $G_N$ is the four-dimensional Newton's constant. After this, Eq. (4.3) can be put in the form of a purely four-dimensional action

\begin{equation}
S_{eff}\supset\int d^4x\frac{\sqrt{-\bar g}{\bar R}}{16\pi G_N}.
\end{equation}

From relationship (4.4) one sees that, as in Ref.\cite{massh}, $G_N$ is a well-defined value even for an infinite extent of the extra dimension (that is the case in our model). We recall that $k$ in the line-element (3.7) does not have the sense of a compactification radius (we have a five-dimensional space-time that is non compact in the extra dimension), it is just a parameter that sets the scale for proper distance measurements along the fifth dimension.

The next step is to determine the physical masses in our theory. It is recommendable to rewrite the matter part of the action (2.6) in the following form

\begin{equation}
S_5^m=\frac{1}{2k}\int_{-\infty}^\infty dy w S^m(y),
\end{equation}
where $w$ is the distribution function (2.5) and $S^m(y)$ is an stochastic four-dimensional action for the matter degrees of freedom living on the fluctuating brane. It is given through the following expression

\begin{equation}
S^m(y)=\int d^4x\sqrt{-g(y)}{\cal L}^m(y),
\end{equation}
where $g_{ab}(y)\equiv G_{ab}(x,y)$ and ${\cal L}^m(y)$ is the Lagrangian of the matter degrees of freedom. The action (4.6) has the sense of some four-dimensional "average action" in respect to the extra dimension. Once explicit integration in $y$ is taken in (4.6), one obtains an effective four-dimensional action. Let us assume that the matter content of the brane is given by a fundamental Higgs field:

\begin{equation}
S^m(y)=\int d^4x\sqrt{-g(y)}[g^{nm}(y)D_n H^\dagger D_m H-\lambda(|H|^2-\upsilon_0^2)^2],
\end{equation}
with a mass parameter $\upsilon_0$. Equation (4.8) can be written in terms of the metric ${\bar g}_{ab}$ as follows

\begin{equation}
S^m(y)={\bar w}\int d^4x\sqrt{-\bar g}{\bar g}^{nm}D_n H^\dagger D_m H-{\bar w}^2\int d^4x\sqrt{-\bar g}\lambda(|H|^2-\upsilon_0^2)^2.
\end{equation}

If we substitute Eq. (4.9) into (4.6) and perform explicit integration we obtain an effective four-dimensional action:

\begin{equation}
S^m_{eff}=\int d^4x\sqrt{-\bar g}[\frac{I_1}{2k}{\bar g}^{nm}D_n H^\dagger D_m H-\frac{I_2}{2k}\lambda(|H|^2-\upsilon_0^2)^2].
\end{equation}
where

\begin{equation}
I_1=\sqrt{\frac{\beta}{2\pi}}\sigma^{-1}\exp[-\frac{(1-\frac{\beta}{4}){\bar y}^2}{2{\bar\sigma}^2}],
\end{equation}
and

\begin{equation}
I_2=\sqrt{\frac{\gamma}{4\pi}}\sigma^{-1}\exp[-\frac{(1-\gamma){\bar y}^2}{{\bar\sigma}^2}],
\end{equation}
with $\beta=[1+(\frac{\bar\sigma}{\sigma})^2]^{-1}$ and $\gamma=[1+\frac{1}{2}(\frac{\bar\sigma}{\sigma})^2]^{-1}$. If we renormalize the wave function $H$, the free-parameter $\lambda$ and the mass parameter $\upsilon_0$ according to the following expressions

\begin{equation}
{\bar H}=(\frac{I_1}{2k})^\frac{1}{2} H,
\end{equation}

\begin{equation}
{\bar\lambda}=2k\frac{I_2}{I_1^2}\lambda,
\end{equation}

\begin{equation}
{\bar\upsilon}_0=(\frac{I_1}{2k})^\frac{1}{2}\upsilon_0,
\end{equation}
then, the effective action (4.10) can be finally written as

\begin{equation}
S^m_{eff}=\int d^4x\sqrt{-\bar g}[{\bar g}^{nm}D_n {\bar H}^\dagger D_m {\bar H}-\bar\lambda(|{\bar H}|^2-{\bar\upsilon}_0^2)^2].
\end{equation}

We see that the physical mass scales are set by the scale ${\bar\upsilon_0}$ in Eq. (4.15). Then, any mass parameter $m_0$ on the fluctuating brane in the five-dimensional theory gives rise to a physical mass

\begin{equation}
{\bar m}_0=(\frac{I_1}{2k})^\frac{1}{2}m_0,
\end{equation}
measured in terms of the metric ${\bar g}_{ab}$. For fixed magnitudes of brane and metric fluctuations $\sigma$ and $\bar\sigma$ in the extra dimension, the desired hierarchy can be achieved by a proper choice of $\bar y$. This is the way in which the mass hierarchy problem is solved "in average" since we performed explicit integration in $y$. However, the mass hierarchy problem finds an alternative resolution in our model if we take a pure stochastic four-dimensional approach.

\subsection{Stochastic approach to the problem}

The model with a brane that attains small fluctuations in the extra dimension is stochastic in nature. We can talk about the probability to find the universe (a four-dimensional hypersurface in the bulk) in a non equilibrium configuration characterized by a non equilibrium value of the external parameter $y\neq 0$, or about the probability to find it in the equilibrium configuration at $y=0$.

We should realize that performing an integration in $y$-coordinate in Eq. (2.6) is "illegal" for a four-dimensional observer (like us) since he does not know anything about the existence of an extra dimension. In this sense the averaging procedure (in respect to the extra dimension) has only a formal meaning, or, it has meaning in respect to a five-dimensional observer living in the bulk space-time. 

The natural way in which a four-dimensional observer can solve the mass hierarchy problem is to forget about integration in $y$ and to work with the stochastic four-dimensional action (4.7). For a fundamental Higgs matter field it takes the form of Eq. (4.8), or, written in terms of the metric ${\bar g}_{ab}$ (that is the metric that appears in the effective Einstein's action (4.5)), it can be put in the form of Eq. (4.9). It is precisely, the main piece for the stochastic approach to the mass hierarchy problem.

Let us to consider a universe in a non equilibrium configuration with $y\neq 0$. Then, if we renormalize the wave function $H$ in Eq. (4.9) as ${\bar H}={\bar w}^\frac{1}{2}H$, the stochastic four-dimensional action (4.9) can be written as

\begin{equation}
S^m(y)=\int d^4x\sqrt{-\bar g}[{\bar g}^{nm}D_n {\bar H}^\dagger D_m {\bar H}-\lambda(|{\bar H}|^2-{\bar w}\upsilon_0^2)^2].
\end{equation}

This implies that, in the pure stochastic approach, the physical mass scales are set by a symmetry-breaking scale

\begin{equation}
{\bar\upsilon}_0={\bar w}^\frac{1}{2}\upsilon_0,
\end{equation}
that depends on position in the extra dimension. As a consequence, any mass parameter $m_0$ in the higher-dimensional theory, will correspond to a physical mass

\begin{equation}
{\bar m}_0=(\sqrt{2\pi}{\bar\sigma})^{-\frac{1}{2}}\exp[-\frac{(y-{\bar y})^2}{4{\bar\sigma}^2}]m_0,
\end{equation}
on the given hypersurface with a non equilibrium configuration. We may argue that the desired mass hierarchy should be generated, however, on the hypersurface with equilibrium configuration at $y=0$. This means that physical masses, on this equilibrium configuration, are given through mass parameters of the five-dimensional theory by the following expression

\begin{equation}
{\bar m}_0=(\sqrt{2\pi}{\bar\sigma})^{-\frac{1}{2}}\exp[-\frac{{\bar y}^2}{4{\bar\sigma}^2}]m_0.
\end{equation}

In consequence, once again, for a fixed value of $\bar\sigma$, we can take account of the desired mass hierarchy by choosing an appropriate value of $\bar y$ (the separation of the centers of the distributions for brane and metric fluctuations in the extra dimension).

In conclusion, the mass hierarchy problem finds a suitable resolution in our model due to a happy circumstance: the brane (together with its matter content) fluctuates about an equilibrium configuration on the hypersurface at $y=0$, while the bound graviton and its lower excited states are confined to a subspace in the neighborhood of $y=\bar y$. Then, as in the RS scenario, the source of the large hierarchy between the observed Planck and weak scales is the small overlapping in the extra dimension of the graviton distribution function $\bar w$ with the universe in the equilibrium configuration (the four-dimensional hypersurface at $y=0$).

\section{Stochastic four-dimensional gravity}

In this section we shall discuss in some detail the stochastic approach to four-dimensional gravity that emerges from our model. Although we are considering small stabilized Gaussian fluctuations in the extra dimension, meaning that the model provides a static description of the five-dimensional space-time, it admits to give some crude cosmological considerations by extrapolating the results obtained in the former sections.

Let us to write the field equations (2.7) in four-dimensional writing, in terms of the metric $\bar g_{ab}$ (Eq. (4.2)). A straightforward manipulation yields

\begin{equation}
\bar R_{ab}-\frac{1}{2}\bar g_{ab}\bar R=8\pi\bar G_N\bar T_{ab}w-\bar\Lambda\bar g_{ab},
\end{equation}
and the constrain equation

\begin{equation}
4\pi\bar G_N w(\bar T-8k^2 T_{44})=\bar\Lambda-\frac{8\pi k^2}{k_5^2\bar\sigma^2}\bar G_N,
\end{equation}
where $\bar R_{ab}$ is the Ricci tensor written in terms of the metric $\bar g_{ab}$, $\bar R=\bar g^{mn}\bar R_{mn}$ and, as in section III,  $\bar T_{ab}=\bar w^{-1}T^5_{ab}$ is the four-dimensional stress-energy tensor of matter coupled to $\bar g_{ab}$ ($\bar T\equiv\bar g^{mn}\bar T_{mn}$). In equations (5.1) and (5.2) we have used the following definitions for the "stochastic" (four-dimensional) Newton's constant

\begin{equation}
\bar G_N\equiv\frac{\bar w k_5^2}{8\pi},
\end{equation}
and the "stochastic" (four-dimensional) cosmological constant

\begin{equation}
\bar\Lambda\equiv 8\pi\bar G_N[V w+\frac{6k^2}{k_5^2\bar\sigma^2}(\frac{(y-\bar y)^2}{\bar\sigma^2}-1)-2\Lambda].
\end{equation}

With the help of equations (5.1) and (5.2), and the definitions (5.3) and (5.4) for "stochastic" constants, one can take account of the stochastic description of four-dimensional gravity that follows from our model. 

From Eq. (5.1) one sees that, if one were to describe Friedmann cosmology (that is not the case in this letter), usual Friedmann behavior would arise. In fact, for the most probable universes to occur ($y=0$), for instance, Eq. (5.1) is just the usual Einstein's equation with fixed values of $\bar G_N(0)=(8\sqrt{2\pi}\pi\bar\sigma)^{-1}\exp[-\frac{\bar y^2}{2\bar\sigma^2}]$ and $\bar\Lambda(0)=8\pi\bar G_N(0)[\frac{V}{\sqrt{2\pi}\sigma}+\frac{6k^2}{k_5^2\bar\sigma^2}(\frac{\bar y^2}{\bar\sigma^2}-1)-2\Lambda]$ respectively. Standard Friedmann behavior is expected in our model, besides, since it contains both ingredients: non null thickness of the "effective" domain where matter degrees of freedom live, and non null bulk component of the five-dimensional stress-energy tensor (see Ref.\cite{kanti}).

The stochastic description provided by Eq. (5.1) is remarkable in another aspect. It admits discussion of the anthropic principle and of the cosmological constant problem. This possibility is linked with the fact that, our model provides distribution functions for the constants appearing in the theory (including the dimensionless gravitational coupling constant $\bar G_N\bar m_0^2=\bar w^2\frac{k_5^2 m_0^2}{8\pi}$ and particles physical masses $\bar m_0$). As an illustration we shall discuss the cosmological constant problem. For this purpose we shall focus on Eq. (5.4). Taking into account Eq. (5.2) and searching for consistency with equations (3.9) and (3.11), Eq. (5.4) can be written as

\begin{equation}
\bar\Lambda=8\pi\bar G_N[\frac{\sqrt{2\pi}\sigma w-1}{\sqrt{2\pi}\sigma} V+\frac{6k^2}{k_5^2\bar\sigma^4}y(y-2\bar y)+\frac{\bar T}{4\sqrt{2\pi}\sigma}].
\end{equation}

Now we shall elaborate on Eq. (5.5). It is a non-Gaussian function in the extra coordinate with a single minimum and two local maxima. The function has two zeros (say at $y_1$ and $y_2$, $y_2>y_1$). For the range $y\in [y_1,y_2]$ it is negative, while it is positive for $y<y_1$ and $y>y_2$. The function (5.5) tends asymptotically to zero at $y\rightarrow\pm\infty$. On the hypersurface located at $y=0$,

\begin{equation}
\bar\Lambda(0)=\frac{\sqrt{2\pi}}{\sigma}\bar G_N (0)\bar T.
\end{equation}

A simple resolution of the cosmological constant problem is reached if we consider that fluctuations of the metric $\bar g_{ab}$ in the extra dimension are very small ($\bar\sigma\sim\sigma$) and,besides, $|\bar y|>>\bar\sigma\sim\sigma$. In this case, for universes "generated" by brane fluctuations about $y=0$, $\bar w=0$ and the cosmological constant is zero for all of them. However from Eq. (5.3) one sees that the four-dimensional Newton's constant is zero as well for all of the universes, in contradiction with the real picture. Therefore, the above assumptions ($\bar\sigma\sim\sigma$ and $\bar y>>\sigma$) are incorrect. Even in the case if one assumes $\bar\sigma\sim\sigma$, one should consider that the centers of the distributions $w$ and $\bar w$ are close enough as to provide the correct hierarchy for $\bar G_N$. In other words, physics requires that there should exist overlapping of the distribution functions $w$ and $\bar w$. The way in which this requirement is realized is not relevant. It should be either by allowing for not so small fluctuations of the metric $\bar g_{ab}$ in the extra dimension ($\bar\sigma>>\sigma$) or by choosing $\bar y\sim\sigma$ or by a combination of both possibilities. In what follows we shall consider the cosmological constant problem in the real situation with overlapping of the distribution functions $w$ and $\bar w$, in such a way that the correct hierarchy for $\bar G_N$ is generated.

In general the four-dimensional cosmological constant $\bar\Lambda$ can take values in the interval $\bar\Lambda\in[\bar\Lambda_{min},\bar\Lambda_{max}]$, where $\bar\Lambda_{min}$ is the minimum (it is a negative value) of the function (5.5) while $\bar\Lambda_{max}$ is its absolute maximum. The range in which $\bar\Lambda$ takes values depends strongly on $\bar\sigma$ and $\bar y$. Therefore, by properly choosing $\bar\sigma$ and $\bar y$, one can drive this range to be as wide as desired.

The following picture emerges. There is a non-null probability to find four-dimensional universes of both, de Sitter ($dS_4$) and anti-de Sitter ($AdS_4$) types, with different magnitudes of the cosmological constant. However, for the most probable universes to occur (those at $y=0$) where the correct hierarchy is generated ($\bar G_N (0)$ is the experimental value of the four-dimensional Newton's constant $\sim 10^{-8}\frac{cm^3}{gr\;s}$), the cosmological constant is given by Eq. (5.6). For a barotropic (perfect) fluid filled universe it can be written as 

\begin{equation}
\bar\Lambda (0)=\bar G_N (0)(3\Gamma-4)\bar\mu,
\end{equation}
where $\bar\mu$ is the renormalized energy density of the fluid ($\bar\mu=\frac{\sqrt{2\pi}}{\sigma}\mu$) and $\Gamma$ is the barotropic index. If we substitute the actual experimental values of $\bar G_N (0)$ and $\bar\mu\sim 10^{-27}\frac{gr}{cm^3}$ (consider the squared speed of light $c^2\approx 10^{21}\frac{cm^2}{s^2}$), Eq. (5.7) yields $\bar\Lambda (0)\sim 10^{-56}cm^{-2}$. We see that, on the four-dimensional hypersurface located at $y=0$ a correct order of magnitude for the cosmological constant is generated.

Since we deal with stabilized brane fluctuations with dispersion $\sigma\sim m_{EW}^{-1}$, in the present approach the cosmological constant problem is not solved. The model just allows for a "coarse" tuning of the cosmological constant instead of a "fine" tuning. However it should be expected that, if one considers a cosmological situation in which the parameters characterizing the distribution evolve in time, the problem can be solved. In fact, if one goes backwards in time, the absolute temperature $T_0$ of the five-dimensional space-time should increase, leading to the width of the distribution $w$ being increased. The Gaussian distribution $w$ is smoothed and tends asymptotically to a constant distribution of probability, meaning that the probability to find a four-dimensional universe with an arbitrary value of the four-dimensional cosmological constant, in the range specified above, is a constant finite magnitude near of the big-bang. When one goes into the future, one sees that the temperature $T_0$ decreases (meaning that $\sigma$ decreases) leading to the distribution of probability being approached by a delta function. This means that irrespective of the initial value of the cosmological constant, for late evolution, the probability to find a universe with a specific value of the cosmological constant is sharply peaked.

\section{The Mach's principle}

It is very encouraging that the stochastic nature of our approach allows for a suitable consideration of the Mach's principle. This principle states, in general, that the local properties of matter are affected, in some way, by the global properties of the universe.

Let us consider, for instance, a particle with physical mass $\bar m_0$ that lives in the universe with an equilibrium configuration. In terms of the mass parameter of the higher-dimensional theory, it is given by the expression (4.21). If we take into account the relationship (3.11), Eq. (4.21) can be written as

\begin{equation}
\bar m_0=(2\pi)^{-\frac{1}{4}}\bar\sigma^{-\frac{1}{2}}\exp[\frac{16k^2T_{44}-4V+16\sqrt{\frac{\pi}{2}}\frac{\sigma}{k^2}\Lambda}{T^4-16k^2T_{44}}]m_0.
\end{equation}

From this equation one sees that the physical mass of the particle (being a local property) depends on the matter content of the space-time, including the brane tension, the bulk cosmological constant and the absolute temperature $T_0$ through $\sigma$. In other words, it depends on global properties of the five-dimensional space-time.

As an illustration, let us consider that fluctuations of the brane in the extra dimension generate an effective five-dimensional stress-energy tensor of matter in the form of a perfect fluid

\begin{equation}
T_{AB}=(\mu+p)u_A u_B+p G_{AB},
\end{equation}
where $\mu$ is the fluid energy density and $p$ is its pressure. Consider, besides, a barotropic equation of state $p=(\Gamma-1)\mu$. Then, in respect to a four-dimensional comoving observer living in the universe at $y=0$, the Eq. (6.1) takes the very simple form

\begin{equation}
\bar m_0=(2\pi)^{-\frac{1}{4}}\bar\sigma^{-\frac{1}{2}}\exp[-4\frac{(\Gamma-1)\mu-V+4\sqrt{\frac{\pi bT_0}{2q_0}}\Lambda}{\Gamma\mu}]m_0.
\end{equation}

It is clear now that the physical mass of any elementary particle depends on such global properties of the universe like its energy density, barotropic index, brane tension, bulk cosmological constant and absolute temperature. We see that the physical mass of the particle (that can be taken as a local property) is very sensitive, in particular, to the state of the fluid. For a dust filled universe ($\Gamma=1$), for instance, it is given by the following expression

\begin{equation}
\bar m_0=(2\pi)^{-\frac{1}{4}}\bar\sigma^{-\frac{1}{2}}\exp[4\frac{V-4\sqrt{\frac{\pi bT_0}{2q_0}}\Lambda}{\mu}]m_0.
\end{equation}

In view of Mach's arguments, it should be expected that, increasing of the energy density of the universe, would yield increasing of the local inertial properties (mass) of the particle. In consequence, from Eq. (6.4) it follows, for positive $V>0$, the following constrain

\begin{equation}
V<4\sqrt{\frac{\pi}{2}}\sigma\Lambda.
\end{equation}

For negative brane tension and $\Gamma\geq 1$, Mach's argument is always fulfilled (see Eq. (6.3)).

Although we used mass to characterize local properties, in the same way we can use other magnitudes characterizing the local properties of the four-dimensional space-time. Take, for instance, the metric (see section III)

\begin{equation}
\bar g_{ab}=\sqrt{2\pi}\bar\sigma e^{\frac{\bar y^2}{2\bar\sigma^2}}g_{ab}.
\end{equation}

From this last expression we see that the local metric properties of the four-dimensional space-time are affected by the matter content of the higher-dimensional space-time structure, the brane tension, the bulk cosmological constant and the absolute temperature (see Eq. (3.11)). In the case studied in this section (a perfect fluid) we see, in agreement with Mach's ideas, that as the energy density of the fluid decreases the components of the metric tensor get increased and, in the zero energy density limit, these are undefined.

\section{Conclusion}

In this letter we have studied a five-dimensional scenario with a non-compact (infinite) extra dimension in which a single three-brane is affected by small Gaussian fluctuations about an equilibrium four-dimensional hypersurface. The model relies on the possibility to consider the bulk space-time as a thermostat with an absolute temperature $T_0$, in which a quasi-isolated subsystem (a three-brane) develops small fluctuations about an equilibrium configuration without breaking the statistical equilibrium in the bulk. Therefore the success of the model depends on the validity of the above considerations. We argued that these could be valid, at least, for particles with energies below the threshold $m_{EW}$, that are confined to the fluctuating brane. In the last instance, thin-brane RS-type scenarios are recovered from the present model in the "no fluctuations" limit ($\sigma\rightarrow 0$) so, it should be taken as an alternative possibility for treating some disturbing issues as the cosmological constant problem and the mass hierarchy.

It is remarkable that, Gaussian fluctuations of the brane, induce Gaussian fluctuations of the metric in the extra dimension with a different characteristic width and a shifted position of the center of the distribution, i. e., two Gaussian distribution functions $w$ and $\bar w$ arise. This happy circumstance allows for an alternative resolution of the mass hierarchy problem. Besides, the RS scenario with two three-branes, one in which the matter degrees of freedom are located and the other in which is confined the graviton, is obtained from the model with a single fluctuating brane, in the no-fluctuations limit.

However, the most remarkable feature of the model is linked with the fact that it provides an intrinsically stochastic description of gravity. In this context, such important issues as the mass hierarchy and cosmological constant problems can be suitably discussed.

It is worthy of mention that the Mach's principle finds a place in our stochastic approach to gravity. It seems that all of then: the Mach's principle, the mass hierarchy and the cosmological constant problem, are different features of a same property of gravity in our model: its stochastic character.

I acknowledge the MES of Cuba for support of this research.


\begin{references}


\bibitem{arkani-hamed} N. Arkani-Hamed, S. Dimopoulos and G. Dvali, Phys. Lett. B{\bf 429}, 263 (1998); I. Antoniadis, N. Arkani-Hamed, S. Dimopoulos and G. Dvali, Phys. Lett. B{\bf 436}, 257 (1998); N. Arkani-Hamed, S. Dimopoulos and G. Dvali, Phys.Rev. D{\bf 59}, 086004 (1999).

\bibitem{randall} L. Randall and R. Sundrum, Nucl. Phys. B{\bf 557}, 79 (1999). 

\bibitem{massh} L. Randall and R. Sundrum, Phys. Rev. Lett. {\bf 83}, 3370 (1999).

\bibitem{sundrum} L. Randall and R. Sundrum, Phys. Rev. Lett. {\bf 83}, 4690 (1999).

\bibitem{lykken} J. Lykken and L. Randall, JHEP {\bf 0006}, 014(2000).

\bibitem{witten} E. Witten, Nucl. Phys. B{\bf 471}, 135 (1996); P. Horava and E. Witten, Nucl. Phys. B{\bf 475}, 94 (1996); J. E. Lidsey, D. Wands and E. J. Copeland, Phys. Rept. {\bf 337}, 343(2000).

\bibitem{deffayet} P. Binetruy, C. Deffayet and D. Langlois, Nucl. Phys. B{\bf 565}, 269 (2000).

\bibitem{kanti} P. Kanti, I. I. Kogan, K. A. Olive and M. Prospelov, Phys. Lett. B{\bf 468}, 31 (1999).

\bibitem{linde} A. S. Goncharov, A. D. Linde and V. F. Mukhanov, Int.J. Mod. Phys. A{\bf 2}, 904 (1988).


\end{references}
\end{document}